\documentclass[journal=jacsat,manuscript=article]{achemso}

\usepackage[version=3]{mhchem} 
\usepackage{graphicx}
\usepackage{dcolumn}
\usepackage{bm}
\usepackage{epstopdf}
\usepackage{subfigure}
\usepackage{amsmath}
\usepackage{color,soul}

\usepackage{tabularx}
\setkeys{acs}{usetitle = true}

\author{Keisuke Takahashi}
\email{keisuke.takahashi@eng.hokudai.ac.jp}
\affiliation{Graduate School of Engineering, Hokkaido University, N-13, W-8, Sapporo 060-8628, Japan}

\author{Shigehito Isobe}
\affiliation{Graduate School of Engineering, Hokkaido University, N-13, W-8, Sapporo 060-8628, Japan}
 
\author{Kengo Omori}
\affiliation{Graduate School of Engineering, Hokkaido University, N-13, W-8, Sapporo 060-8628, Japan}

\author{Torge Mashoff}
\affiliation{Center for Nanotechnology Innovation at NEST, Istituto Italiano di Tecnologia, Piazza San Silvestro 12, 56127 Pisa, Italy}

\author{Domenica Convertino}
\affiliation{Center for Nanotechnology Innovation at NEST, Istituto Italiano di Tecnologia, Piazza San Silvestro 12, 56127 Pisa, Italy}

\author{Vaidotas Miseikis}
\affiliation{Center for Nanotechnology Innovation at NEST, Istituto Italiano di Tecnologia, Piazza San Silvestro 12, 56127 Pisa, Italy}

\author{Camilla Coletti}
\affiliation{Center for Nanotechnology Innovation at NEST, Istituto Italiano di Tecnologia, Piazza San Silvestro 12, 56127 Pisa, Italy}

\author{Valentina Tozzini}
\affiliation{NEST, Istituto Nanoscienze-CNR and Scuola Normale Superiore, Piazza San Silvestro 12, 56127 Pisa, Italy}

\author{Stefan Heun}
\email{stefan.heun@nano.cnr.it}
\affiliation{NEST, Istituto Nanoscienze-CNR and Scuola Normale Superiore, Piazza San Silvestro 12, 56127 Pisa, Italy}

\title{Revealing the multi-bonding state between hydrogen and graphene--sup\-port\-ed Ti clusters}
\abbreviations{IR,NMR,UV}
\keywords{American Chemical Society, \LaTeX}

\begin{document}
\begin{abstract}

Hydrogen adsorption on graphene--sup\-port\-ed metal clusters has brought much controversy due to the complex nature of the bonding between hydrogen and metal clusters. 
The bond types of hydrogen and graphene--sup\-port\-ed Ti clusters are experimentally and theoretically investigated. 
Transmission electron microscopy shows that Ti clusters of nanometer--size are formed on graphene. 
Thermal desorption spectroscopy captures three hydrogen desorption peaks from hydrogenated graphene--sup\-port\-ed Ti clusters. 
First principle calculations also found  three types of interaction: Two types of bonds with different partial ionic character and physisorption. The physical origin for this rests on the charge state of the Ti clusters: when Ti clusters are neutral, H$_2$ is dissociated, and H forms bonds with the Ti cluster. On the other hand, H$_2$ is adsorbed in molecular form on positively charged Ti clusters, resulting in physisorption. Thus, this work clarifies the bonding mechanisms of hydrogen on graphene--sup\-port\-ed Ti clusters.

\end{abstract}

\section{Introduction}
\label{1}
The unique nature of metal clusters and two--di\-men\-sion\-al materials have advanced the field of materials science; however, a fundamental understanding behind scientific phenomena related to such materials often leads to further questions \cite{mingos1990introduction,geim2007rise}. 
Utilizing metal clusters and two--di\-men\-sion\-al materials results in functional materials that possess outstanding physical and chemical properties \cite{stankovich2006graphene,pumera2011graphene, C1CS15078B}. 
Two--di\-men\-sion\-al materials like graphene are found to be good substrates for supporting metal clusters due to their chemical stability and high surface to volume ratio \cite{krasheninnikov2009embedding,Mashoff2013,Takahashi2014}. 
Graphene--sup\-port\-ed metal clusters show high reactivity which leads towards applications in the fields of catalysis and energy storage \cite{liang2011co3o4,wu2012graphene,guo2012fept}.

The chemistry of metal clusters on graphene involves complex bonding between metal clusters and gas molecules when gases such as hydrogen are introduced, resulting in difficulty of experimental measurements and analysis. 
In particular, hydrogen adsorption on graphene--sup\-port\-ed metal clusters often leads to controversial arguments over how hydrogen is adsorbed, whether the hydrogen adsorption takes place on the metal clusters or on defect sites of graphene, whether spillover effects occur, and what types of bonds between hydrogen and metal are formed. 
Experimental studies report multiple hydrogen desorption peaks, and the hydrogen desorption properties are difficult to reproduce \cite{parambhath2011investigation,parambhath2012effect,hudson2014hydrogen,Mashoff2013,Takahashi2014}.

In order to achieve a fundamental understanding of the interactions between hydrogen and graphene--sup\-port\-ed metal clusters, first principle calculations and experimental measurements are performed. 
In particular, Ti clusters when combined with graphene are predicted to be a particularly suitable material for hydrogen adsorption with range of 3.6~wt$\%$ to 7.8~wt$\%$ hydrogen uptake \cite{Durgun2008,Bhattacharya2010,liu2010titanium,valencia2015trends,ramos2016ti4}. 
Therefore, in this paper the interactions between hydrogen and graphene--sup\-port\-ed Ti clusters are theoretically and experimentally explored. 

\section{Materials and methods}
SiC and Cu substrates are implemented in order to support graphene, since such substrates are reported to preserve the electronic structure of graphene even in the presence of metal clusters \cite{takahashi2015growth}.

Graphene for transmission electron microscopy (TEM) observation was synthesized using the chemical vapor deposition method  \cite{li2011large}.
The graphene was then transferred to a Cu--based TEM grid. 
Ti clusters were deposited on graphene using a vacuum deposition technique where the size of clusters is controlled by substrate temperature and deposition time.
An aberration--corrected TEM (FEI, TITAN) was used for observing the Ti clusters on a graphene/TEM grid. The acceleration voltage was set to 60 kV. 

Hydrogenation and dehydrogenation of graphene--sup\-port\-ed Ti clusters was performed in an ultra--high vacuum chamber with a base pressure of $5 \times 10^{-11}$ mbar. Graphene for hydrogenation and dehydrogenation analysis was synthesized on SiC(0001) which was used for scanning tunneling microscopy (STM) observation in previous work \cite{Mashoff2013,Goler2013a,Goler2013,Mashoff2015}. Before titanium deposition, graphene samples were annealed at 900K for several hours to remove adsorbents and to obtain a clean surface. This was done by direct current heating to ensure a homogeneous sample temperature. The high quality of the pristine graphene films was verified by atomically resolved STM images.\cite{Mashoff2013,Mashoff2015} Titanium was deposited on graphene at room temperature using a commercial electron--beam evaporator. The Ti--coverage was calibrated by STM imaging. The amount of depositied Ti for the samples discussed here was 0.84 monolayers (1 monolayer (ML) = $1.32 \times 10^{15}$ atoms/cm$^2$).\cite{Mashoff2015} All temperatures were measured using a thermocouple mounted on the sample holder, directly in contact with the sample and additionally cross--calibrated with a pyrometer.

Hydrogenation of Ti clusters on graphene for thermal desorption spectroscopy (TDS) measurements was accomplished by exposing them to molecular deuterium for 5 minutes at a pressure of $1 \times 10^{-7}$~mbar at 95~K. Deuterium (D$_2$, mass 4) was used instead of hydrogen (H$_2$, mass 2) for a better signal--to--noise ratio in TDS. Deuterium is chemically identical to hydrogen; however, note that the desorption temperatures might be slightly shifted owing to the well known isotope effect \cite{Gdowski1986,Zecho2002}. For the TDS measurements the samples were positioned in front of a mass spectrometer and heated at a constant rate of 10~K/s up to a temperature of approximately 1000~K, while recording the mass 4--channel of the mass spectrometer. The measured TDS signal was cross--calibrated by the read--out of the pressure gauge in the same vacuum chamber, using an ion gauge sensitivity factor of 0.35.\cite{Helden2012}

First principle calculations employing the Grid based projector augmented wave method (GPAW) are implemented for electronic structure analysis of hydrogen and Ti clusters on graphene \cite{mortensen2005real}.
The exchange-correlation vdW-DF functional, namely PBE\cite{pbe} functional corrected  with DF correction within the Grimme-like\cite{grimme} scheme is used in order to account for the van der Waals force between the Cu substrate and graphene, and physisorption of hydrogen \cite{dion2004van}. 

Grid spacing is set to 0.20 \AA\ with 0.1 eV of smearing and spin polarization calculations.
Spin polarization calculation is applied for all calculations.
$(4 \times 4 \times 1)$ of special k points of the Brillouin zone sampling used within periodic boundary conditions and 15 \AA\ of vaccum is applied to the z axis \cite{monkhorst1976special}. Bader charge analysis is implemented for calculating the electron transfers \cite{sanville2007improved,henkelman2006fast}.

The model system consists of a supercell of Cu(111)(2x2) with four atomic layers of Cu where the bottom two layers of Cu are fixed and a single layer of graphene (consisting of 32 C atoms) located on top. 
Different model systems with Ti clusters of various size are then build based on the TEM images and increasingly hydrogenated (see Results section). A global optimization scheme, employing the Basin--hopping algorithm, is implemented in order to find the ground state structures of Ti clusters on graphene/Cu(111) \cite{Takahashi2013}. 

Hydrogen adsorption/desorption energies (E$_{ad}$) are calculated as:
\begin{equation}
   E_{ad}=-(E[Sur+H{_2}]-E[Sur]-E[H{_2}]).
   \label{eq:eq1}
\end{equation}
Here $E[Sur]$ represents the energy of graphene--sup\-port\-ed Ti clusters while E[H$_2$] is the energy of gas phase H$_2$. Note that a positive sign indicates an exothermic reaction.

The binding energies (E$_{b}$) between graphene and Ti clusters are calculated as:
\begin{equation}
   E_{b}=-\frac{(E[Ti+Sub]-E[Sub]-nE[Ti_{n}])}{n}.
   \label{eq:eq2}
\end{equation}

Cohesive energies(E$_{co}$) for supported Ti clusters are calculated as:
\begin{equation}
   E_{co}=-\frac{(E[Ti+Sub]-E[Sub]-nE[Ti_{1}])}{n}.
   \label{eq:eq3}
\end{equation}
Here $E[Sub]$  represents the energy of the graphene substrate, and n is the number of Ti atoms.

\section{Results}
\begin{figure}[ht]
  \includegraphics[width=85 mm]{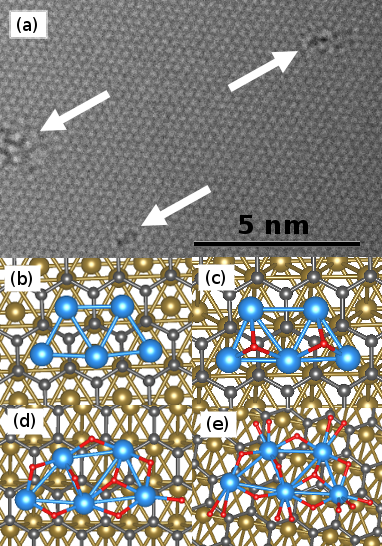}
  \caption{\label{fig:fig1}(a) Transmission electron microscopy image of Ti clusters on single layer graphene. Ti clusters are indicated by white arrows. (b--e) Selected structure models of Ti clusters on graphene: (b) Ti$_5$, (c)Ti$_5$:1H$_2$, (d)Ti$_5$:4H$_2$, (e)Ti$_5$:9H$_2$. Color code: C--black, Cu--brown, H--red, Ti--blue.}
\end{figure}

A TEM image of Ti clusters on single-layer graphene on Cu is shown in Fig.~\ref{fig:fig1}(a). 
The figure shows an atomically--resolved single layer graphene film on which four Ti clusters are indicated by white arrows, with two small Ti clusters being pointed at by the middle arrow. The size of each Ti cluster is found to be 1.37 nm, 0.27nm, 0.68 nm, and 1.0 nm, from left to right. 
The important observation here is that Ti is not distributed as individual atoms on the surface, as often assumed in theoretical investigations \cite{Durgun2008,Kim2009,Bhattacharya2010,Lee2010,Liu2010,Chu2011,Fair2013}, but due to a high cohesive energy \cite{Bhattacharya2010,Lee2010,Fair2013} it forms clusters. Note that nanometer--sized Ti clusters have previously also been observed on graphene on SiC \cite{Mashoff2013,Mashoff2015}.

{\color{black}
Based on the size of the Ti clusters in Fig.~\ref{fig:fig1}(a), atomic models of Ti$_n$ $(n=1-5)$ clusters on graphene are constructed and optimized.
The calculations suggest that Ti clusters grow with a trapezoidal shape which is found to have the lowest energy, as depicted in Fig.~\ref{fig:fig1}(b). 
The size of a Ti$_5$ cluster is calculated to be 0.53~nm. 
The magnetic moment of graphene supported Ti$_n$ clusters $(n=1-5)$ is calculated to be $1.49\mu{_B}$, $0.59\mu{_B}$, $0.79\mu{_B}$, $0.27\mu{_B}$, and $0.13\mu{_B}$, respectively, which is smaller than that of Ti clusters in the gas phase.\cite{medina2010structural}
The binding energies between graphene/Cu and Ti clusters and the cohesive energies of supported Ti clusters are calculated and shown in Table \ref{tab:table1}. In particular, Ti clusters are adsorbed on the graphene/Cu substrate with relatively high stability. Similar phenomena have also been reported for the case of graphene supported Pd clusters \cite{ramos2015palladium,Johll2011}.}

\begin{table*}[ht]
\caption{\label{tab:table1}The binding energies (E$_{b}$) between graphene/Cu and Ti clusters and the cohesive energies (E$_{co}$) of supported Ti clusters per atom in eV.
}
\begin{tabular}{lcc}
\hline
 & E$_{b}$ & E$_{co}$ \\ 
\hline
Ti$_1$ & 5.27 & N/A \\
Ti$_2$ & 4.22 & 5.07  \\
Ti$_3$ & 3.66 & 5.09 \\
Ti$_4$ & 3.08 & 4.85  \\
Ti$_5$ & 2.82 & 4.68  \\
\hline
\end{tabular}
\end{table*}
\begin{figure}[ht]
  \includegraphics[width=85 mm]{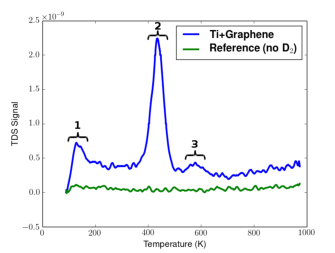}
  \caption{\label{fig:fig2}TDS spectrum of hydrogen desorption from Ti clusters sup\-port\-ed on graphene. The reference curve was obtained from a sample which was not exposed to molecular deuterium.}
\end{figure}

Hydrogen desorption analysis of hydrogenated Ti clusters sup\-port\-ed on graphene is investigated using thermal desorption spectroscopy (TDS). The resulting temperature-dependent desorption curve is shown in Fig.~\ref{fig:fig2}. The TDS spectrum presents three hydrogen desorption peaks. 
In particular, major hydrogen release occurs at 420~K while there are two more hydrogen desorption peaks at 150K and 580K. 
In a control experiment, the sample was not exposed to a deuterium flux. The resulting desorption curve did not show any desorption peaks.

The amount of desorbed (and therefore stored) deuterium can be estimated from these TDS data. Subtraction of the background and integration of the TDS curve gives the area under the curve, $F = 3.7 \times 10^8$ mbar $\cdot$ s. At a given pressure, the amount of desorbed gas equals the pumping speed of the vacuum system, here $S = 375$ l/s. This leads to the amount of desorbed deuterium $pV = FS = nRT$, with $R = 8.314$ J~K$^{-1}$ mol$^{-1}$ the gas constant, resulting in $n = 5.6 \times 10^{-10}$ mol = $3.4 \times 10^{14}$ desorbed D$_2$ molecules. This quantity can be used to evaluate the Gravimetric Density ($GD$) of the system, i.e., the ratio of loaded hydrogen mass over total system mass (from now on, we consider hydrogen and not deuterium). The general formula for $GD$ can then be written as
\begin{equation}
	GD = \frac{M_H}{M_{Ti} + M_G + M_H},\label{eq:eq4}
\end{equation}
with $M_n$ the mass of titanium ($n = Ti$), graphene ($n = G$), and hydrogen ($n = H$). $3.4 \times 10^{14}$~H$_2$ molecules have a mass of $M_H = 1.1 \times 10^{-12}$ kg. The sample has a size of 12.4~mm$^2$, and graphene a mass density of $7.6 \times 10^{-7}$ kg/m$^2$,\cite{Chen2008} which results in a mass of the graphene of $M_G = 9.4 \times 10^{-12}$ kg. On the same sample area, 0.84 ML of Ti correspond to $1.4 \times 10^{14}$ atoms, corresponding to a weight of $M_{Ti} = 1.1 \times 10^{-11}$ kg. Using eq. \ref{eq:eq4}, the $GD$ finally results to be $GD = 5.1$~wt\%. Furthermore, by comparing the numbers of released hydrogen molecules and Ti surface atoms, it can be deduced that each Ti atom in average can bind 2.4 hydrogen molecules.

The approximate desorption energy barriers $E_d$ from the measured desorption temperatures $T_d$ were then estimated. First-order desorption is assumed, based on the reported molecule-like dimer arrangement of hydrogen
atoms on graphene,\cite{Goler2013} and similar to what has been observed for hydrogen release from graphite\cite{Denisov2001} and Rh(110).\cite{Vesselli2008} Defining $\tau_m$ as the time from the start of the desorption ramp to the moment at which the desorption temperature $T_d$ is reached, $\tau_m$ equals $(T_d - T_s) / \beta$ with $T_s$ the temperature at the start of the ramp (here $T_s = 95$~K) and $\beta$ the heating rate (here $\beta = 10$~K/s). Then one has $E_d / k_B T_d = A \tau_m e^{-E_d / k_B T_d}$,\cite{Woodruff1994} with $A$ the Arrhenius constant (typical value $10^{13}$~s$^{-1}$). For $T_d = 150$~K $\tau_m = 6$~s and $E_d / k_B T_d =28.4$ were obtained; hence $E_d = 0.37$~eV/molecule. This desorption peak can therefore be classified as related to physisorption ($E_d < 1.0$~eV/molecule). On the other hand, for the desorption peaks at 420~K and 580~K $E_d = 1.1$~eV/molecule or 0.55 eV/atom and $E_d = 1.5$~eV/molecule = 0.75 eV/atom were obtained, respectively, suggesting these peaks are related to chemisorption ($E_d > 1.0$~eV/molecule). In using the Redhead equation,\cite{Woodruff1994,Mashoff2013} the same values for the desorption energy barriers were obtained. These estimates of the relation between desorption temperature and desorption barrier are also consistent with other values reported for hydrogen desorption from various materials.\cite{Helden2012,Goler2013}

These multiple hydrogen release peaks address a controversial issue relating to the hydrogen adsorption sites. 
The location of such sites is often debated to be either on the graphene side or the metal side \cite{parambhath2011investigation,parambhath2012effect,hudson2014hydrogen,Mashoff2013,Takahashi2014}. 
Calculations provide controversial predictions, suggesting that multi--bonding states between hydrogen and Ti clusters might be involved. 

In order to reveal the details of the interaction between hydrogen and Ti clusters on graphene, hydrogen adsorption over Ti$_n$ $(n=1-5)$ clusters on graphene is simulated. 
Hydrogen adsorption at various adsorption sites is considered: on clean graphene, on graphene with a defect site, on graphene with a defect site which is filled by a Ti$_1$ cluster, and on Ti$_n$ $(n=1-5)$ clusters on graphene. 
Hydrogen adsorption is performed by adding molecular H$_2$ in a stepwise manner until the Ti$_n$ clusters are fully hydrogenated.

\begin{figure}[ht]
  \includegraphics[width=85 mm]{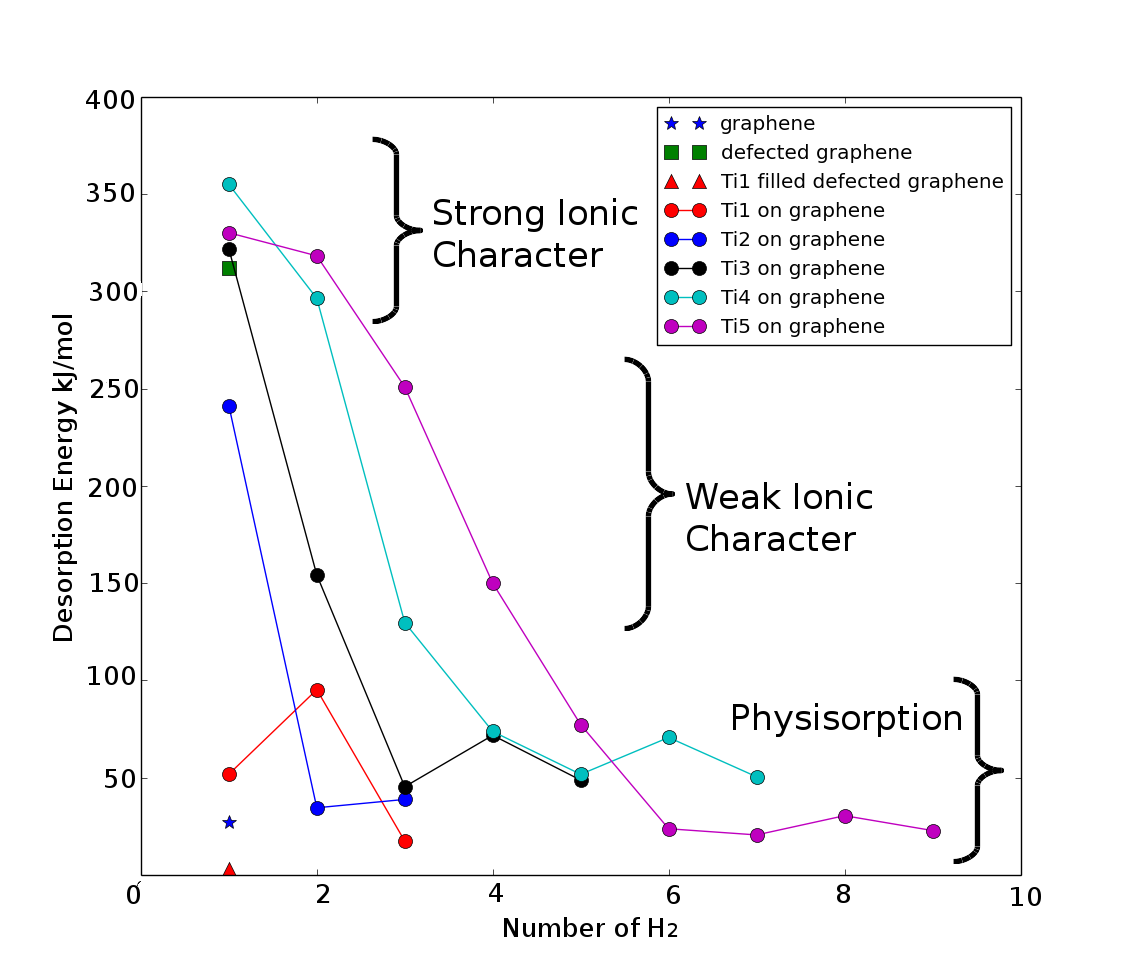}
  \caption{\label{fig:fig3}Calculated desorption energies for hydrogen adsorbed on clean graphene, defect site of graphene, defect site of graphene filled with Ti$_1$ cluster, and Ti$_n$ $(n=1-5)$ clusters on graphene.
  Strong ionic character and weak ionic character represent chemisorption while physisorption is mediated by van der Waals force.}
\end{figure}

As shown in Fig.~\ref{fig:fig3}, the hydrogen desorption energy on clean graphene is calculated to be 27 kJ/mol (or 0.28 eV/H$_2$), which indicates that hydrogen is adsorbed on graphene by physisorption. 
On the other hand, hydrogen adsorption at the defect in graphene has a relatively high desorption energy of 313 kJ/mol. 
Here H$_2$ is dissociated and adsorbed as H atoms at the edge of the defect. 
These cases represent the two different bond types, namely physisorption and chemisorption, that occur during hydrogen and graphene interaction, where physisorption is based on weak van der Waals forces between the H$_2$ molecule and graphene, while chemisorption is based on C--H chemical bonds that form after H$_2$ molecule dissociation.
Conversely, the hydrogen desorption energy from a Ti$_1$ cluster occupying a defect site in graphene is significantly low at 3 kJ/mol. 
This could be attributed to the strong bond between Ti$_1$ and the defect site of graphene mediated by a charge transfer.

According to the first principle calculations, hydrogen adsorption on Ti$_1$ clusters on graphene is governed by physisorption as no charge transfer is involved between H$_2$ molecules and Ti. 
This is in agreement with previous work showing that up to four H$_2$ molecules can be stored by single graphene--sup\-port\-ed Ti atoms \cite{Rojas2007,valencia2015trends}. 
However, larger clusters Ti$_n$ $(n=2-5)$ behave differently. Ti$_2$ induces dissociative adsorption of the first H$_2$ with 0.5 electrons transfered from Ti$_2$ to each H atom. 
This behavior, previously observed in isolated small Ti clusters \cite{Shang2009}, can be described as a polarized covalent bond with partial ionic character. 
The second and third H$_2$ are weakly adsorbed in molecular form  where electron transfer between Ti and H$_2$ is not involved.
A similar behavior is also observed for hydrogen adsorption on Ti$_n$ $(n=3-5)$.

Structural details of the different bond types between hydrogen and Ti$_5$ are shown in Figs.~\ref{fig:fig1}(c--e). Chemical bonding is seen in Ti$_5$:1H$_2$ where H$_2$ is dissociated and adsorbed as H atoms, as shown in Fig.~\ref{fig:fig1}(c). 
A weaker bonding is observed in  Ti$_5$:4H$_2$. 
Though the molecules are fully dissociated, the charge transfer and ionic character slightly decrease, and the binding energy is lower. 
As more H$_2$ molecules are added, the physical character of the interaction prevails, forming a shell of physisorbed molecules as in Ti$_5$:9H$_2$ shown in Fig.~\ref{fig:fig1}(e).

Bader charge analysis provides further information for an understanding of the multi-bonding between hydrogen and Ti clusters. 
Two regimes of charge transfer from H to Ti are identified, namely "strong" in the range of $0.6-0.7$ electrons/H and "weak" with $0.4-0.5$ electrons/H; for physisorption, a charge transfer is not involved. 
In addition, Bader analysis reveals interesting results in hydrogen adsorption at the defected site of the graphene where H is positively charged by 0.5 electron which is the opposite charge state of hydrogen on Ti clusters. 
As the size of the Ti clusters grows, not only can more hydrogen be adsorbed but also the number of hydrogen atoms that fit in each type of bond increases. 
This can be seen in Fig.~\ref{fig:fig3}: a Ti$_5$ cluster, for example, can store up to 9 hydrogen molecules.

First principle calculations reveals that positively charged Ti clusters are not able to dissociate H$_2$, leading to a  weak H$_2$ adsorption through physisorption, while neutral Ti clusters tend to transfer their electrons towards H$_2$, resulting in H$_2$ dissociation and chemical bonding of H.  
Similar results were reported for Ti clusters adsorbed on fullerenes.\cite{sun2005clustering} 
One can conclude that the bond type between H$_2$ and Ti clusters is strongly dependent on the charge state of the Ti clusters. 
In particular, initially introduced H$_2$ tends to dissociate and be adsorbed through chemical bonding due to the large charge transfer from the Ti clusters. 
As the number of adsorbed H$_2$ on the Ti cluster increases, the strength and the ionic character of the bonds between H$_2$ and the Ti cluster weakens. 
Once Ti clusters are fully positively charged, H$_2$ is then adsorbed in molecular form via physisorption. 
As a result, three bonding regions are both observed in experiment and calculations as shown in Figures \ref{fig:fig2} and \ref{fig:fig3}, respectively. 
Thus, the link between experimental hydrogen desorption shown in Fig.~\ref{fig:fig2} and three different types of bonding between hydrogen and Ti clusters based on first principle calculations shown in Fig.~\ref{fig:fig3} is established.

In addition,  the gravimetric density of hydrogen is evaluated where hydrogen storage capacity is estimated to be 5.1\%, corresponding to 2.4 hydrogen molecules for each Ti atom, in line with previous studies of Ti functionalized graphene on SiC\cite{Mashoff2015} 
and above the values obtained for non-functionalized graphene based materials in same conditions ($\sim$1\% at RT).\cite{alex} 
A rough estimation of the corresponding volumetric density leads to an optimal value $\sim$0.2~kg/l, which would allow interesting automotive applications. 
This however implies the capability of building 3D frameworks with controllable inter-layer spacing or porosity (a value of $\sim$ 1nm is assumed in the estimate). 
Therefore, future developments of these studies for H-storage applications shall involve not only the optimization of the Ti clusters concentration and distribution, 
but also the chemical functionalization of the sheets aimed at building 3D frameworks with controlled structural properties.

\section{Conclusions}

In conclusion, the bonding mechanisms between hydrogen and graphene--sup\-port\-ed Ti clusters are experimentally and theoretically investigated. Ti clusters are synthesized on graphene using vacuum deposition techniques. 
TEM confirms that nanometer--sized Ti clusters are formed on graphene. 
TDS shows that three hydrogen desorption peaks from hydrogenated Ti clusters occur. 
First principle calculations reveal three bond types: chemical bonding with strong ionic character, weak chemical bonding with smaller ionic character, and physisorption. 
In particular, we show that the physical origin for this rests on the charge state of the Ti clusters. 
When Ti clusters are neutral, H$_2$ is dissociated and adsorbed through ionic bonding, while H$_2$ is adsorbed in molecular form over positively charged Ti clusters.

\section{Acknowledgment}

CPU time is funded by the Japan Society for the Promotion of Science and is performed at Hokkaido University, Sapporo, Japan. 
This work has been partially supported by ENEOS Hydrogen Trust Fund. 
Funding from the European Union Seventh Framework Programme under Grant Agreement No. 696656 Graphene Core1 is acknowledged. 
Financial support from the CNR in the framework of the agreements on scientific collaboration between CNR and JSPS (Japan), CNRS (France), NRF (Korea), and RFBR (Russia) is acknowledged.
SH acknowledges funding from the Italian Ministry of Foreign Affairs, Direzione Generale per la Promozione del Sistema Paese (agreements on scientific collaboration with Canada (Quebec) and Poland).

\begin{figure}[ht]
  \includegraphics[width=85 mm]{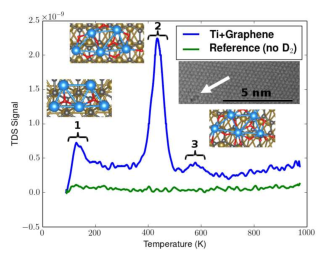}
  \caption{\label{fig:fig4}Table of Content.}
\end{figure}





\providecommand{\latin}[1]{#1}
\providecommand*\mcitethebibliography{\thebibliography}
\csname @ifundefined\endcsname{endmcitethebibliography}
  {\let\endmcitethebibliography\endthebibliography}{}

\newpage
 "For Table of Contents Use Only,"

\begin{figure}[h]
  \centering        
  {\includegraphics[width=85 mm]{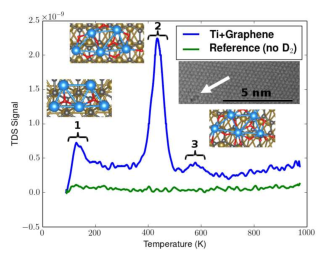}}
  \caption*{Table of Contents}
\end{figure}






\begin{mcitethebibliography}{52}
\providecommand*\natexlab[1]{#1}
\providecommand*\mciteSetBstSublistMode[1]{}
\providecommand*\mciteSetBstMaxWidthForm[2]{}
\providecommand*\mciteBstWouldAddEndPuncttrue
  {\def\EndOfBibitem{\unskip.}}
\providecommand*\mciteBstWouldAddEndPunctfalse
  {\let\EndOfBibitem\relax}
\providecommand*\mciteSetBstMidEndSepPunct[3]{}
\providecommand*\mciteSetBstSublistLabelBeginEnd[3]{}
\providecommand*\EndOfBibitem{}
\mciteSetBstSublistMode{f}
\mciteSetBstMaxWidthForm{subitem}{(\alph{mcitesubitemcount})}
\mciteSetBstSublistLabelBeginEnd
  {\mcitemaxwidthsubitemform\space}
  {\relax}
  {\relax}

\bibitem[Mingos and Wales(1990)Mingos, and Wales]{mingos1990introduction}
Mingos,~D. M.~P.; Wales,~D.~J. \emph{{Introduction to Cluster Chemistry}};
  Prentice Hall, 1990\relax
\mciteBstWouldAddEndPuncttrue
\mciteSetBstMidEndSepPunct{\mcitedefaultmidpunct}
{\mcitedefaultendpunct}{\mcitedefaultseppunct}\relax
\EndOfBibitem
\bibitem[Geim and Novoselov(2007)Geim, and Novoselov]{geim2007rise}
Geim,~A.~K.; Novoselov,~K.~S. {The Rise of Graphene}. \emph{Nature Materials}
  \textbf{2007}, \emph{6}, 183--191\relax
\mciteBstWouldAddEndPuncttrue
\mciteSetBstMidEndSepPunct{\mcitedefaultmidpunct}
{\mcitedefaultendpunct}{\mcitedefaultseppunct}\relax
\EndOfBibitem
\bibitem[Stankovich \latin{et~al.}(2006)Stankovich, Dikin, Dommett, Kohlhaas,
  Zimney, Stach, Piner, Nguyen, and Ruoff]{stankovich2006graphene}
Stankovich,~S.; Dikin,~D.~A.; Dommett,~G.~H.; Kohlhaas,~K.~M.; Zimney,~E.~J.;
  Stach,~E.~A.; Piner,~R.~D.; Nguyen,~S.~T.; Ruoff,~R.~S. {Graphene-based
  Composite Materials}. \emph{Nature} \textbf{2006}, \emph{442}, 282--286\relax
\mciteBstWouldAddEndPuncttrue
\mciteSetBstMidEndSepPunct{\mcitedefaultmidpunct}
{\mcitedefaultendpunct}{\mcitedefaultseppunct}\relax
\EndOfBibitem
\bibitem[Pumera(2011)]{pumera2011graphene}
Pumera,~M. {Graphene-based nanomaterials for energy storage}. \emph{Energy \&
  Environmental Science} \textbf{2011}, \emph{4}, 668--674\relax
\mciteBstWouldAddEndPuncttrue
\mciteSetBstMidEndSepPunct{\mcitedefaultmidpunct}
{\mcitedefaultendpunct}{\mcitedefaultseppunct}\relax
\EndOfBibitem
\bibitem[Huang \latin{et~al.}(2012)Huang, Qi, Boey, and Zhang]{C1CS15078B}
Huang,~X.; Qi,~X.; Boey,~F.; Zhang,~H. {Graphene-based composites}. \emph{Chem.
  Soc. Rev.} \textbf{2012}, \emph{41}, 666--686\relax
\mciteBstWouldAddEndPuncttrue
\mciteSetBstMidEndSepPunct{\mcitedefaultmidpunct}
{\mcitedefaultendpunct}{\mcitedefaultseppunct}\relax
\EndOfBibitem
\bibitem[Krasheninnikov \latin{et~al.}(2009)Krasheninnikov, Lehtinen, Foster,
  Pyykk{\"o}, and Nieminen]{krasheninnikov2009embedding}
Krasheninnikov,~A.; Lehtinen,~P.; Foster,~A.; Pyykk{\"o},~P.; Nieminen,~R.
  {Embedding Transition-Metal Atoms in Graphene: Structure, Bonding, and
  Magnetism}. \emph{Phys. Rev. Lett.} \textbf{2009}, \emph{102}, 126807\relax
\mciteBstWouldAddEndPuncttrue
\mciteSetBstMidEndSepPunct{\mcitedefaultmidpunct}
{\mcitedefaultendpunct}{\mcitedefaultseppunct}\relax
\EndOfBibitem
\bibitem[Mashoff \latin{et~al.}(2013)Mashoff, Takamura, Tanabe, Hibino,
  Beltram, and Heun]{Mashoff2013}
Mashoff,~T.; Takamura,~M.; Tanabe,~S.; Hibino,~H.; Beltram,~F.; Heun,~S.
  {Hydrogen Storage with Titanium-Functionalized Graphene}. \emph{Appl. Phys.
  Lett.} \textbf{2013}, \emph{103}, 013903\relax
\mciteBstWouldAddEndPuncttrue
\mciteSetBstMidEndSepPunct{\mcitedefaultmidpunct}
{\mcitedefaultendpunct}{\mcitedefaultseppunct}\relax
\EndOfBibitem
\bibitem[Takahashi \latin{et~al.}(2014)Takahashi, Wang, Chiba, Nakagawa, Isobe,
  and Ohnuki]{Takahashi2014}
Takahashi,~K.; Wang,~Y.; Chiba,~S.; Nakagawa,~Y.; Isobe,~S.; Ohnuki,~S. {Low
  Temperature Hydrogenation of Iron Nanoparticles on Graphene}. \emph{Sci.
  Rep.} \textbf{2014}, \emph{4}, 4598\relax
\mciteBstWouldAddEndPuncttrue
\mciteSetBstMidEndSepPunct{\mcitedefaultmidpunct}
{\mcitedefaultendpunct}{\mcitedefaultseppunct}\relax
\EndOfBibitem
\bibitem[Liang \latin{et~al.}(2011)Liang, Li, Wang, Zhou, Wang, Regier, and
  Dai]{liang2011co3o4}
Liang,~Y.; Li,~Y.; Wang,~H.; Zhou,~J.; Wang,~J.; Regier,~T.; Dai,~H.
  {Co$_3$O$_4$ Nanocrystals on Graphene as a Synergistic Catalyst for Oxygen
  Reduction Reaction}. \emph{Nat. Mater.} \textbf{2011}, \emph{10},
  780--786\relax
\mciteBstWouldAddEndPuncttrue
\mciteSetBstMidEndSepPunct{\mcitedefaultmidpunct}
{\mcitedefaultendpunct}{\mcitedefaultseppunct}\relax
\EndOfBibitem
\bibitem[Wu \latin{et~al.}(2012)Wu, Zhou, Yin, Ren, Li, and
  Cheng]{wu2012graphene}
Wu,~Z.-S.; Zhou,~G.; Yin,~L.-C.; Ren,~W.; Li,~F.; Cheng,~H.-M. {Graphene/metal
  oxide composite electrode materials for energy storage}. \emph{Nano Energy}
  \textbf{2012}, \emph{1}, 107--131\relax
\mciteBstWouldAddEndPuncttrue
\mciteSetBstMidEndSepPunct{\mcitedefaultmidpunct}
{\mcitedefaultendpunct}{\mcitedefaultseppunct}\relax
\EndOfBibitem
\bibitem[Guo and Sun(2012)Guo, and Sun]{guo2012fept}
Guo,~S.; Sun,~S. {FePt nanoparticles assembled on graphene as enhanced catalyst
  for oxygen reduction reaction}. \emph{Journal of the American Chemical
  Society} \textbf{2012}, \emph{134}, 2492--2495\relax
\mciteBstWouldAddEndPuncttrue
\mciteSetBstMidEndSepPunct{\mcitedefaultmidpunct}
{\mcitedefaultendpunct}{\mcitedefaultseppunct}\relax
\EndOfBibitem
\bibitem[Parambhath \latin{et~al.}(2011)Parambhath, Nagar, Sethupathi, and
  Ramaprabhu]{parambhath2011investigation}
Parambhath,~V.~B.; Nagar,~R.; Sethupathi,~K.; Ramaprabhu,~S. {Investigation of
  Spillover Mechanism in Palladium Decorated Hydrogen Exfoliated Functionalized
  Graphene}. \emph{J. Phys. Chem. C} \textbf{2011}, \emph{115},
  15679--15685\relax
\mciteBstWouldAddEndPuncttrue
\mciteSetBstMidEndSepPunct{\mcitedefaultmidpunct}
{\mcitedefaultendpunct}{\mcitedefaultseppunct}\relax
\EndOfBibitem
\bibitem[Parambhath \latin{et~al.}(2012)Parambhath, Nagar, and
  Ramaprabhu]{parambhath2012effect}
Parambhath,~V.~B.; Nagar,~R.; Ramaprabhu,~S. {Effect of nitrogen doping on
  hydrogen storage capacity of palladium decorated graphene}. \emph{Langmuir}
  \textbf{2012}, \emph{28}, 7826--7833\relax
\mciteBstWouldAddEndPuncttrue
\mciteSetBstMidEndSepPunct{\mcitedefaultmidpunct}
{\mcitedefaultendpunct}{\mcitedefaultseppunct}\relax
\EndOfBibitem
\bibitem[Hudson \latin{et~al.}(2014)Hudson, Raghubanshi, Awasthi, Sadhasivam,
  Bhatnager, Simizu, Sankar, and Srivastava]{hudson2014hydrogen}
Hudson,~M. S.~L.; Raghubanshi,~H.; Awasthi,~S.; Sadhasivam,~T.; Bhatnager,~A.;
  Simizu,~S.; Sankar,~S.; Srivastava,~O. {Hydrogen Uptake of Reduced Graphene
  Oxide and Graphene Sheets Decorated with Fe Nanoclusters}. \emph{Int. J.
  Hydrogen Energy} \textbf{2014}, \emph{39}, 8311--8320\relax
\mciteBstWouldAddEndPuncttrue
\mciteSetBstMidEndSepPunct{\mcitedefaultmidpunct}
{\mcitedefaultendpunct}{\mcitedefaultseppunct}\relax
\EndOfBibitem
\bibitem[Durgun \latin{et~al.}(2008)Durgun, Ciraci, and Yildirim]{Durgun2008}
Durgun,~E.; Ciraci,~S.; Yildirim,~T. {Functionalization of Carbon-based
  Nanostructures with Light Transition-Metal Atoms for Hydrogen Storage}.
  \emph{Phys. Rev. B} \textbf{2008}, \emph{77}, 085405\relax
\mciteBstWouldAddEndPuncttrue
\mciteSetBstMidEndSepPunct{\mcitedefaultmidpunct}
{\mcitedefaultendpunct}{\mcitedefaultseppunct}\relax
\EndOfBibitem
\bibitem[Bhattacharya \latin{et~al.}(2010)Bhattacharya, Bhattacharya, Majumder,
  and Das]{Bhattacharya2010}
Bhattacharya,~A.; Bhattacharya,~S.; Majumder,~C.; Das,~G.~P. {Transition-Metal
  Decoration Enhanced Room-Temperature Hydrogen Storage in a Defect-Modulated
  Graphene Sheet}. \emph{J. Phys. Chem. C} \textbf{2010}, \emph{114},
  10297\relax
\mciteBstWouldAddEndPuncttrue
\mciteSetBstMidEndSepPunct{\mcitedefaultmidpunct}
{\mcitedefaultendpunct}{\mcitedefaultseppunct}\relax
\EndOfBibitem
\bibitem[Liu \latin{et~al.}(2010)Liu, Ren, He, and Cheng]{liu2010titanium}
Liu,~Y.; Ren,~L.; He,~Y.; Cheng,~H.-P. {Titanium-decorated Graphene for
  High-capacity Hydrogen Storage Studied by Density Functional Simulations}.
  \emph{J. Phys. Condens. Matter} \textbf{2010}, \emph{22}, 445301\relax
\mciteBstWouldAddEndPuncttrue
\mciteSetBstMidEndSepPunct{\mcitedefaultmidpunct}
{\mcitedefaultendpunct}{\mcitedefaultseppunct}\relax
\EndOfBibitem
\bibitem[Valencia \latin{et~al.}(2015)Valencia, Gil, and
  Frapper]{valencia2015trends}
Valencia,~H.; Gil,~A.; Frapper,~G. Trends in the Hydrogen Activation and
  Storage by Adsorbed 3d Transition Metal Atoms onto Graphene and Nanotube
  Surfaces: A DFT Study and Molecular Orbital Analysis. \emph{J Phys Chem C}
  \textbf{2015}, \emph{119}, 5506\relax
\mciteBstWouldAddEndPuncttrue
\mciteSetBstMidEndSepPunct{\mcitedefaultmidpunct}
{\mcitedefaultendpunct}{\mcitedefaultseppunct}\relax
\EndOfBibitem
\bibitem[Ramos-Castillo \latin{et~al.}(2016)Ramos-Castillo, Reveles,
  Cifuentes-Quintal, Zope, and de~Coss]{ramos2016ti4}
Ramos-Castillo,~C.~M.; Reveles,~J.~U.; Cifuentes-Quintal,~M.~E.; Zope,~R.~R.;
  de~Coss,~R. Ti$_4$-- and Ni$_4$--Doped Defective Graphene Nanoplatelets as
  Efficient Materials for Hydrogen Storage. \emph{J. Phys. Chem. C}
  \textbf{2016}, \emph{120}, 5001\relax
\mciteBstWouldAddEndPuncttrue
\mciteSetBstMidEndSepPunct{\mcitedefaultmidpunct}
{\mcitedefaultendpunct}{\mcitedefaultseppunct}\relax
\EndOfBibitem
\bibitem[Takahashi(2015)]{takahashi2015growth}
Takahashi,~K. {The growth of Fe clusters over graphene/Cu (111)}. \emph{2D
  Materials} \textbf{2015}, \emph{2}, 014001\relax
\mciteBstWouldAddEndPuncttrue
\mciteSetBstMidEndSepPunct{\mcitedefaultmidpunct}
{\mcitedefaultendpunct}{\mcitedefaultseppunct}\relax
\EndOfBibitem
\bibitem[Li \latin{et~al.}(2011)Li, Magnuson, Venugopal, Tromp, Hannon, Vogel,
  Colombo, and Ruoff]{li2011large}
Li,~X.; Magnuson,~C.~W.; Venugopal,~A.; Tromp,~R.~M.; Hannon,~J.~B.;
  Vogel,~E.~M.; Colombo,~L.; Ruoff,~R.~S. {Large-Area Graphene Single Crystals
  Grown by Low-Pressure Chemical Vapor Deposition of Methane on Copper}.
  \emph{J. Am. Chem. Soc.} \textbf{2011}, \emph{133}, 2816--2819\relax
\mciteBstWouldAddEndPuncttrue
\mciteSetBstMidEndSepPunct{\mcitedefaultmidpunct}
{\mcitedefaultendpunct}{\mcitedefaultseppunct}\relax
\EndOfBibitem
\bibitem[Goler \latin{et~al.}(2013)Goler, Coletti, Piazza, Pingue, Colangelo,
  Pellegrini, Emtsev, Forti, Starke, Beltram, and Heun]{Goler2013a}
Goler,~S.; Coletti,~C.; Piazza,~V.; Pingue,~P.; Colangelo,~F.; Pellegrini,~V.;
  Emtsev,~K.~V.; Forti,~S.; Starke,~U.; Beltram,~F.; Heun,~S. Revealing the
  atomic structure of the buffer layer between SiC(0001) and epitaxial
  graphene. \emph{Carbon} \textbf{2013}, \emph{51}, 249\relax
\mciteBstWouldAddEndPuncttrue
\mciteSetBstMidEndSepPunct{\mcitedefaultmidpunct}
{\mcitedefaultendpunct}{\mcitedefaultseppunct}\relax
\EndOfBibitem
\bibitem[Goler \latin{et~al.}(2013)Goler, Coletti, Tozzini, Piazza, Mashoff,
  Beltram, Pellegrini, and Heun]{Goler2013}
Goler,~S.; Coletti,~C.; Tozzini,~V.; Piazza,~V.; Mashoff,~T.; Beltram,~F.;
  Pellegrini,~V.; Heun,~S. Influence of Graphene Curvature on Hydrogen
  Adsorption: Toward Hydrogen Storage Devices. \emph{J. Phys. Chem. C}
  \textbf{2013}, \emph{117}, 11506\relax
\mciteBstWouldAddEndPuncttrue
\mciteSetBstMidEndSepPunct{\mcitedefaultmidpunct}
{\mcitedefaultendpunct}{\mcitedefaultseppunct}\relax
\EndOfBibitem
\bibitem[Mashoff \latin{et~al.}(2015)Mashoff, Convertino, Miseikis, Coletti,
  Piazza, Tozzini, Beltram, and Heun]{Mashoff2015}
Mashoff,~T.; Convertino,~D.; Miseikis,~V.; Coletti,~C.; Piazza,~V.;
  Tozzini,~V.; Beltram,~F.; Heun,~S. {Increasing the Active Surface of Titanium
  Islands on Graphene by Nitrogen Sputtering}. \emph{Appl. Phys. Lett.}
  \textbf{2015}, \emph{106}, 083901\relax
\mciteBstWouldAddEndPuncttrue
\mciteSetBstMidEndSepPunct{\mcitedefaultmidpunct}
{\mcitedefaultendpunct}{\mcitedefaultseppunct}\relax
\EndOfBibitem
\bibitem[Gdowski and Felter(1986)Gdowski, and Felter]{Gdowski1986}
Gdowski,~G.~E.; Felter,~T.~E. {Summary Abstract: An Observed Isotope Effect for
  Hydrogen and Deuterium Adsorption/Desorption on Pd(111)}. \emph{J. Vac. Sci.
  Technol. A} \textbf{1986}, \emph{4}, 1409\relax
\mciteBstWouldAddEndPuncttrue
\mciteSetBstMidEndSepPunct{\mcitedefaultmidpunct}
{\mcitedefaultendpunct}{\mcitedefaultseppunct}\relax
\EndOfBibitem
\bibitem[Zecho \latin{et~al.}(2002)Zecho, G{\"u}ttler, Sha, Jackson, and
  K{\"u}ppers]{Zecho2002}
Zecho,~T.; G{\"u}ttler,~A.; Sha,~X.; Jackson,~B.; K{\"u}ppers,~J. {Adsorption
  of Hydrogen and Deuterium Atoms on the (0001) Graphite Surface}. \emph{J.
  Chem. Phys.} \textbf{2002}, \emph{117}, 8486\relax
\mciteBstWouldAddEndPuncttrue
\mciteSetBstMidEndSepPunct{\mcitedefaultmidpunct}
{\mcitedefaultendpunct}{\mcitedefaultseppunct}\relax
\EndOfBibitem
\bibitem[van Helden \latin{et~al.}(2012)van Helden, van~den Berg, and
  Weststrate]{Helden2012}
van Helden,~P.; van~den Berg,~J.-A.; Weststrate,~C.~J. Hydrogen Adsorption on
  Co Surfaces: A Density Functional Theory and Temperature Programmed
  Desorption Study. \emph{ACS Catal.} \textbf{2012}, \emph{2}, 1097\relax
\mciteBstWouldAddEndPuncttrue
\mciteSetBstMidEndSepPunct{\mcitedefaultmidpunct}
{\mcitedefaultendpunct}{\mcitedefaultseppunct}\relax
\EndOfBibitem
\bibitem[Mortensen \latin{et~al.}(2005)Mortensen, Hansen, and
  Jacobsen]{mortensen2005real}
Mortensen,~J.; Hansen,~L.; Jacobsen,~K. {Real-space Grid Implementation of the
  Projector Augmented Wave Method}. \emph{Phys. Rev. B} \textbf{2005},
  \emph{71}, 035109\relax
\mciteBstWouldAddEndPuncttrue
\mciteSetBstMidEndSepPunct{\mcitedefaultmidpunct}
{\mcitedefaultendpunct}{\mcitedefaultseppunct}\relax
\EndOfBibitem
\bibitem[Perdew \latin{et~al.}(1996)Perdew, Burke, and Ernzerhof]{pbe}
Perdew,~J.~P.; Burke,~K.; Ernzerhof,~M. Generalized Gradient Approximation Made
  Simple. \emph{Phys. Rev. Lett.} \textbf{1996}, \emph{77}, 3865--3868\relax
\mciteBstWouldAddEndPuncttrue
\mciteSetBstMidEndSepPunct{\mcitedefaultmidpunct}
{\mcitedefaultendpunct}{\mcitedefaultseppunct}\relax
\EndOfBibitem
\bibitem[Grimme(2006)]{grimme}
Grimme,~S. Semiempirical GGA-type Density Functional Constructed with a
  Long-Range Dispersion Correction. \emph{J. Comput. Chem.} \textbf{2006},
  \emph{27}, 1787\relax
\mciteBstWouldAddEndPuncttrue
\mciteSetBstMidEndSepPunct{\mcitedefaultmidpunct}
{\mcitedefaultendpunct}{\mcitedefaultseppunct}\relax
\EndOfBibitem
\bibitem[Dion \latin{et~al.}(2004)Dion, Rydberg, Schr{\"o}der, Langreth, and
  Lundqvist]{dion2004van}
Dion,~M.; Rydberg,~H.; Schr{\"o}der,~E.; Langreth,~D.~C.; Lundqvist,~B.~I. {Van
  der Waals Density Functional for General Geometries}. \emph{Phys. Rev. Lett.}
  \textbf{2004}, \emph{92}, 246401\relax
\mciteBstWouldAddEndPuncttrue
\mciteSetBstMidEndSepPunct{\mcitedefaultmidpunct}
{\mcitedefaultendpunct}{\mcitedefaultseppunct}\relax
\EndOfBibitem
\bibitem[Monkhorst and Pack(1976)Monkhorst, and Pack]{monkhorst1976special}
Monkhorst,~H.~J.; Pack,~J.~D. {Special points for Brillouin-zone Integrations}.
  \emph{Phys. Rev. B} \textbf{1976}, \emph{13}, 5188--5192\relax
\mciteBstWouldAddEndPuncttrue
\mciteSetBstMidEndSepPunct{\mcitedefaultmidpunct}
{\mcitedefaultendpunct}{\mcitedefaultseppunct}\relax
\EndOfBibitem
\bibitem[Sanville \latin{et~al.}(2007)Sanville, Kenny, Smith, and
  Henkelman]{sanville2007improved}
Sanville,~E.; Kenny,~S.~D.; Smith,~R.; Henkelman,~G. {Improved Grid-based
  Algorithm for Bader Charge Allocation}. \emph{J. Comput. Chem.}
  \textbf{2007}, \emph{28}, 899--908\relax
\mciteBstWouldAddEndPuncttrue
\mciteSetBstMidEndSepPunct{\mcitedefaultmidpunct}
{\mcitedefaultendpunct}{\mcitedefaultseppunct}\relax
\EndOfBibitem
\bibitem[Henkelman \latin{et~al.}(2006)Henkelman, Arnaldsson, and
  J{\'o}nsson]{henkelman2006fast}
Henkelman,~G.; Arnaldsson,~A.; J{\'o}nsson,~H. {A Fast and Robust Algorithm for
  Bader Decomposition of Charge Density}. \emph{Comput. Mater. Sci.}
  \textbf{2006}, \emph{36}, 354--360\relax
\mciteBstWouldAddEndPuncttrue
\mciteSetBstMidEndSepPunct{\mcitedefaultmidpunct}
{\mcitedefaultendpunct}{\mcitedefaultseppunct}\relax
\EndOfBibitem
\bibitem[Takahashi \latin{et~al.}(2013)Takahashi, Isobe, and
  Ohnuki]{Takahashi2013}
Takahashi,~K.; Isobe,~S.; Ohnuki,~S. {Chemisorption of Hydrogen on Fe Clusters
  Through Hybrid Bonding Mechanisms}. \emph{Appl. Phys. Lett.} \textbf{2013},
  \emph{102}, 113108\relax
\mciteBstWouldAddEndPuncttrue
\mciteSetBstMidEndSepPunct{\mcitedefaultmidpunct}
{\mcitedefaultendpunct}{\mcitedefaultseppunct}\relax
\EndOfBibitem
\bibitem[Kim \latin{et~al.}(2009)Kim, Jhi, Lim, and Park]{Kim2009}
Kim,~G.; Jhi,~S.-H.; Lim,~S.; Park,~N. {Effect of Vacancy Defects in Graphene
  on Metal Anchoring and Hydrogen Adsorption}. \emph{Appl. Phys. Lett.}
  \textbf{2009}, \emph{94}, 173102\relax
\mciteBstWouldAddEndPuncttrue
\mciteSetBstMidEndSepPunct{\mcitedefaultmidpunct}
{\mcitedefaultendpunct}{\mcitedefaultseppunct}\relax
\EndOfBibitem
\bibitem[Lee \latin{et~al.}(2010)Lee, Ihm, Cohen, and Louie]{Lee2010}
Lee,~H.; Ihm,~J.; Cohen,~M.~L.; Louie,~S.~G. {Calcium-Decorated Graphene-Based
  Nanostructures for Hydrogen Storage}. \emph{Nano Lett.} \textbf{2010},
  \emph{10}, 793\relax
\mciteBstWouldAddEndPuncttrue
\mciteSetBstMidEndSepPunct{\mcitedefaultmidpunct}
{\mcitedefaultendpunct}{\mcitedefaultseppunct}\relax
\EndOfBibitem
\bibitem[Liu \latin{et~al.}(2010)Liu, Ren, He, and Cheng]{Liu2010}
Liu,~Y.; Ren,~L.; He,~Y.; Cheng,~H.-P. {Titanium-decorated Graphene for
  High-capacity Hydrogen Storage Studied by Density Functional Simulations}.
  \emph{J. Phys.: Condens. Matter} \textbf{2010}, \emph{22}, 445301\relax
\mciteBstWouldAddEndPuncttrue
\mciteSetBstMidEndSepPunct{\mcitedefaultmidpunct}
{\mcitedefaultendpunct}{\mcitedefaultseppunct}\relax
\EndOfBibitem
\bibitem[Chu \latin{et~al.}(2011)Chu, Hu, Hu, Yang, and Deng]{Chu2011}
Chu,~S.; Hu,~L.; Hu,~X.; Yang,~M.; Deng,~J. {Titanium-Embedded Graphene as
  High-Capacity Hydrogen-Storage Media}. \emph{Int. J. Hydrogen Energy}
  \textbf{2011}, \emph{36}, 12324\relax
\mciteBstWouldAddEndPuncttrue
\mciteSetBstMidEndSepPunct{\mcitedefaultmidpunct}
{\mcitedefaultendpunct}{\mcitedefaultseppunct}\relax
\EndOfBibitem
\bibitem[Fair \latin{et~al.}(2013)Fair, Cui, Li, Shieh, Zheng, Liu, Delley,
  Ford, Ringer, and Stampfl]{Fair2013}
Fair,~K.~M.; Cui,~X.~Y.; Li,~L.; Shieh,~C.~C.; Zheng,~R.~K.; Liu,~Z.~W.;
  Delley,~B.; Ford,~M.~J.; Ringer,~S.~P.; Stampfl,~C. {Hydrogen Adsorption
  Capacity of Adatoms on Double Carbon Vacancies of Graphene: A Trend Study
  from First Principles}. \emph{Phys. Rev. B} \textbf{2013}, \emph{87},
  014102\relax
\mciteBstWouldAddEndPuncttrue
\mciteSetBstMidEndSepPunct{\mcitedefaultmidpunct}
{\mcitedefaultendpunct}{\mcitedefaultseppunct}\relax
\EndOfBibitem
\bibitem[Medina \latin{et~al.}(2010)Medina, {de Coss}, Tapia, and
  Canto]{medina2010structural}
Medina,~J.; {de Coss},~R.; Tapia,~A.; Canto,~G. {Structural, Energetic and
  Magnetic Properties of Small Ti$_n$ (n= 2--13) Clusters: A Density Functional
  Study}. \emph{Eur. Phys. J. B} \textbf{2010}, \emph{76}, 427\relax
\mciteBstWouldAddEndPuncttrue
\mciteSetBstMidEndSepPunct{\mcitedefaultmidpunct}
{\mcitedefaultendpunct}{\mcitedefaultseppunct}\relax
\EndOfBibitem
\bibitem[Ramos-Castillo \latin{et~al.}(2015)Ramos-Castillo, Reveles, Zope, and
  de~Coss]{ramos2015palladium}
Ramos-Castillo,~C.; Reveles,~J.; Zope,~R.; de~Coss,~R. {Palladium Clusters
  Supported on Graphene Monovacancies for Hydrogen Storage}. \emph{J. Phys.
  Chem. C} \textbf{2015}, \emph{119}, 8402\relax
\mciteBstWouldAddEndPuncttrue
\mciteSetBstMidEndSepPunct{\mcitedefaultmidpunct}
{\mcitedefaultendpunct}{\mcitedefaultseppunct}\relax
\EndOfBibitem
\bibitem[Johll \latin{et~al.}(2011)Johll, Wu, Ong, Kang, and Tok]{Johll2011}
Johll,~H.; Wu,~J.; Ong,~S.~W.; Kang,~H.~C.; Tok,~E.~S. Graphene-adsorbed Fe,
  Co, and Ni trimers and tetramers: Structure, stability, and magnetic moment.
  \emph{Phys. Rev. B} \textbf{2011}, \emph{83}, 205408\relax
\mciteBstWouldAddEndPuncttrue
\mciteSetBstMidEndSepPunct{\mcitedefaultmidpunct}
{\mcitedefaultendpunct}{\mcitedefaultseppunct}\relax
\EndOfBibitem
\bibitem[Chen \latin{et~al.}(2008)Chen, Jang, Xiao, Ishigami, and
  Fuhrer]{Chen2008}
Chen,~J.-H.; Jang,~C.; Xiao,~S.; Ishigami,~M.; Fuhrer,~M.~S. Intrinsic and
  extrinsic performance limits of graphene devices on SiO$_2$. \emph{Nat.
  Nanotech.} \textbf{2008}, \emph{3}, 206\relax
\mciteBstWouldAddEndPuncttrue
\mciteSetBstMidEndSepPunct{\mcitedefaultmidpunct}
{\mcitedefaultendpunct}{\mcitedefaultseppunct}\relax
\EndOfBibitem
\bibitem[Denisov and Kompaniets(2001)Denisov, and Kompaniets]{Denisov2001}
Denisov,~E.~A.; Kompaniets,~T.~N. Kinetics of Hydrogen Release from Graphite
  after Hydrogen Atom Sorption. \emph{Phys. Scr.} \textbf{2001}, \emph{T94},
  128\relax
\mciteBstWouldAddEndPuncttrue
\mciteSetBstMidEndSepPunct{\mcitedefaultmidpunct}
{\mcitedefaultendpunct}{\mcitedefaultseppunct}\relax
\EndOfBibitem
\bibitem[Vesselli \latin{et~al.}(2008)Vesselli, Campaniello, Baraldi,
  Bianchettin, Africh, Esch, Lizzit, and Comelli]{Vesselli2008}
Vesselli,~E.; Campaniello,~M.; Baraldi,~A.; Bianchettin,~L.; Africh,~C.;
  Esch,~F.; Lizzit,~S.; Comelli,~G.~A. Surface Core Level Shift Study of
  Hydrogen-Induced Ordered Structures on Rh(110). \emph{J. Phys. Chem. C}
  \textbf{2008}, \emph{112}, 14475\relax
\mciteBstWouldAddEndPuncttrue
\mciteSetBstMidEndSepPunct{\mcitedefaultmidpunct}
{\mcitedefaultendpunct}{\mcitedefaultseppunct}\relax
\EndOfBibitem
\bibitem[Woodruff and Delchar(1994)Woodruff, and Delchar]{Woodruff1994}
Woodruff,~D.~P.; Delchar,~T.~A. \emph{Modern Techniques of Surface Science};
  Cambridge University Press: New York, 1994\relax
\mciteBstWouldAddEndPuncttrue
\mciteSetBstMidEndSepPunct{\mcitedefaultmidpunct}
{\mcitedefaultendpunct}{\mcitedefaultseppunct}\relax
\EndOfBibitem
\bibitem[Rojas and Leiva(2007)Rojas, and Leiva]{Rojas2007}
Rojas,~M.~I.; Leiva,~E. P.~M. {Density Functional Theory Study of a Graphene
  Sheet Modified with Titanium in Contact with Different Adsorbates}.
  \emph{Phys. Rev. B} \textbf{2007}, \emph{76}, 155415\relax
\mciteBstWouldAddEndPuncttrue
\mciteSetBstMidEndSepPunct{\mcitedefaultmidpunct}
{\mcitedefaultendpunct}{\mcitedefaultseppunct}\relax
\EndOfBibitem
\bibitem[Shang \latin{et~al.}(2009)Shang, Wei, and Zhu]{Shang2009}
Shang,~M.-H.; Wei,~S.-H.; Zhu,~Y.-J. {The Evolution of Geometric and Electronic
  Structures for the Hydrogen Storage on Small Ti$_n$ $(n = 2-7)$ Clusters}.
  \emph{J. Phys. Chem. C} \textbf{2009}, \emph{113}, 15507\relax
\mciteBstWouldAddEndPuncttrue
\mciteSetBstMidEndSepPunct{\mcitedefaultmidpunct}
{\mcitedefaultendpunct}{\mcitedefaultseppunct}\relax
\EndOfBibitem
\bibitem[Sun \latin{et~al.}(2005)Sun, Wang, Jena, and
  Kawazoe]{sun2005clustering}
Sun,~Q.; Wang,~Q.; Jena,~P.; Kawazoe,~Y. {Clustering of Ti on a C60 Surface and
  Its Effect on Hydrogen Storage}. \emph{J. Am. Chem. Soc.} \textbf{2005},
  \emph{127}, 14582--14583\relax
\mciteBstWouldAddEndPuncttrue
\mciteSetBstMidEndSepPunct{\mcitedefaultmidpunct}
{\mcitedefaultendpunct}{\mcitedefaultseppunct}\relax
\EndOfBibitem
\bibitem[Klechikov \latin{et~al.}(2015)Klechikov, Mercier, Merino, Blanco,
  Merino, and Talyzin]{alex}
Klechikov,~A.~G.; Mercier,~G.; Merino,~P.; Blanco,~S.; Merino,~C.;
  Talyzin,~A.~V. Hydrogen storage in bulk graphene-related materials.
  \emph{Micr. Mes. Mater.} \textbf{2015}, \emph{210}, 46--51\relax
\mciteBstWouldAddEndPuncttrue
\mciteSetBstMidEndSepPunct{\mcitedefaultmidpunct}
{\mcitedefaultendpunct}{\mcitedefaultseppunct}\relax
\EndOfBibitem
\end{mcitethebibliography}
\end{document}